\documentclass[twocolumn,superscriptaddress,secnumarabic,
amssymb,amsmath,nobibnotes,aps,prd,showpacs,nofootinbib]{revtex4}%
\usepackage{graphicx}
\usepackage{epsf}
\usepackage{bm}
\usepackage{amsmath}
\usepackage{amsfonts}
\usepackage{amssymb}
\usepackage{epstopdf}
\usepackage{natbib}%
\usepackage{rotating}
\usepackage{pdflscape}
\usepackage{adjustbox}
\providecommand{\U}[1]{\protect\rule{.1in}{.1in}}
\newcommand{\be}{\begin{equation}}
\newcommand{\ee}{\end{equation}}

\newcommand{\mincir}{\raise
-3.truept\hbox{\rlap{\hbox{$\sim$}}\raise4.truept\hbox{$<$}\ }}
\newcommand{\magcir}{\raise
-3.truept\hbox{\rlap{\hbox{$\sim$}}\raise4.truept\hbox{$>$}\ }}

\usepackage{color}
\usepackage{changes}

\begin{document}

\title{Dark calling Dark: Interaction in the dark sector in presence of neutrino properties after Planck CMB final release}

\author{Weiqiang Yang}
\email{d11102004@163.com}
\affiliation{Department of Physics, Liaoning Normal University, Dalian, 116029, People's Republic of China}

\author{Supriya Pan}
\email{supriya.maths@presiuniv.ac.in}
\affiliation{Department of Mathematics, Presidency University, 86/1 College Street, Kolkata 700073, India}

\author{Rafael C. Nunes}
\email{rafadcnunes@gmail.com}
\affiliation{Divis\~{a}o de Astrof\'{i}sica, Instituto Nacional de Pesquisas Espaciais, Avenida dos Astronautas 1758, S\~{a}o Jos\'{e} dos Campos, 12227-010, SP, Brazil}

\author{David F. Mota}
\email{d.f.mota@astro.uio.no}
\affiliation{Institute of Theoretical Astrophysics, University of Oslo, P.O. Box 1029 Blindern, N-0315 Oslo, Norway}

\begin{abstract}
We investigate a well known scenario of interaction in the dark sector where the vacuum energy is interacting with cold dark matter throughout the cosmic evolution 
in light of the cosmic microwave background (CMB) data from final Planck 2018 release. In addition to this minimal scenario, we generalize the model baseline by including the properties of neutrinos, such as the neutrino mass scale ($M_{\nu}$) and the effective number of neutrino species ($N_{\rm eff}$) as free parameters, in order to verify the possible effects that such parameters might generate on the coupling parameter, and vice versa. As already known, we again confirm that in light of the Planck 2018 data, such dark coupling can successfully solve the $H_0$ tension (with and without the presence of neutrinos). Concerning the properties of neutrinos, we find that $M_{\nu}$ may be wider than expected within the $\Lambda$CDM model and $N_{\rm eff}$ is fully compatible with three neutrino species (similar to $\Lambda$CDM prevision). The parameters characterizing the properties of neutrinos do not correlate with the coupling parameter of the interaction model. When considering the joint analysis of CMB from Planck 2018 and an estimate of $H_0$ from Hubble Space Telescope 2019 data, {\it we find an evidence for a non-null value of the coupling parameter at more than 3$\sigma$ confidence-level.} We also discuss the possible effects on the interacting scenario due to the inclusion of baryon acoustic oscillations  data with 
Planck  2018. Our main results updating the dark sectors' interaction and neutrino properties in the model baseline, represent a new perspective in this direction. Clearly, a possible new physics in light of some dark interaction between dark energy and dark matter can serve as an alternative to $\Lambda$CDM scenario to explain the observable Universe, mainly in light of the current tension on $H_0$. 
\end{abstract}

\maketitle
\section{Introduction}
\label{sec:intro}

Within the context of General theory of Relativity the observational signatures of our Universe are well described by two dark components, namely the dark matter (DM), responsible for the structure formation of our Universe and an exotic dark fluid having high negative pressure, known as dark energy (DE)  driving the expansion of our Universe in an accelerated way. The observational data from various independent astronomical sources predict that nearly 68\% of the total energy budget of the Universe is occupied by DE and about 28\% of the total budget of the universe is formed by DM \cite{Ade:2015xua,Aghanim:2018eyx}. The remaining 4\% is contributed by non-relativistic baryons, relativistic radiation, neutrinos and other particles. This standard description of our Universe  is well described within the well known $\Lambda$CDM paradigm.  However, even after a series of independent observations, 
the intrinsic nature of DM and DE, are still indeterminate. Thus, over the past several years a number of cosmological models with different variants have been introduced and tested by many investigators.  Out of such models, the possibility of a non-gravitational interaction between the dark components got massive attention for explaining possible observational deficiencies presented in the standard $\Lambda$CDM scenario, see an incomplete list of works on this specific area  \cite{Amendola:1999er,Cai:2004dk,Pavon:2005yx,Barrow:2006hia,Amendola:2006dg,Koivisto:2006ai,He:2008tn,Valiviita:2008iv,delCampo:2008sr,delCampo:2008jx,Gavela:2009cy,Majerotto:2009np,Clemson:2011an,Pan:2013rha,Yang:2014vza,Yang:2014gza,Nunes:2014qoa,Faraoni:2014vra,Pan:2012ki,Yang:2014hea,Shahalam:2015sja,Li:2015vla,Nunes:2016dlj,Yang:2016evp,Pan:2016ngu,Mukherjee:2016shl,Sharov:2017iue,Shahalam:2017fqt,Guo:2017hea,Cai:2017yww,Yang:2017yme,Yang:2017ccc,Yang:2017zjs,Pan:2017ent,Yang:2018pej,Yang:2018ubt,Paliathanasis:2019hbi,Pan:2019jqh,Yang:2019bpr,Yang:2019vni,Yang:2019uzo} (also see two recent review articles in this direction \cite{Bolotin:2013jpa,Wang:2016lxa}). However, the strongest support in favor of interaction between the dark sectors appears from some recent observational results indicating  for a possible non-null interaction in the dark sector \cite{Salvatelli:2014zta,Kumar:2016zpg,Kumar:2017dnp,DiValentino:2017iww,Yang:2018euj,Pan:2019gop}. Additionally, it has been observed in recent years that some phenomenological interaction models in the dark sector may ease the tensions on some important cosmological parameters, such as the tension on $H_0$ \cite{Kumar:2016zpg,Kumar:2017dnp,DiValentino:2017iww,Yang:2018euj,Yang:2018uae,Kumar:2019wfs,Pan:2019gop,DiValentino:2019ffd} and the tension on $S_8$ \cite{Kumar:2017dnp,Kumar:2019wfs,Kumar_Nunes:2017,Pourtsidou:2016ico,An:2017crg}. Although it is very difficult to ease the tension on $H_0$ and $S_8$ simultaneously, but a recent investigation  \cite{Kumar:2019wfs} reports that such possibility is feasible if a non-gravitational coupling in the dark sector is allowed.  A direct estimation on this possible dark interaction was quantified in \cite{Kumar_Nunes:2017}, with scattering cross section $\leq 10^{-29} \,\, {\rm cm}^2$, for typical DM mass scale. Such observational indications are not yet conclusive and details at theoretical, observational and statistical level still need to be investigated to support a possible observational preference for a dark coupling.

On the other hand, neutrino properties play a very crucial role in the dynamics of our Universe, by inferring the direct changes on some  important cosmological sources, and consequently, in the determination of cosmological parameters (see an incomplete list of recent and past works in this direction \cite{Lesgourgues:2006nd,Lattanzi:2017ubx,LaVacca:2009yp,Giusarma:2016phn,Gerbino:2016sgw,Bellomo:2016xhl,Nunes:2017xon,Vagnozzi:2017ovm,Yang:2017amu,Vagnozzi:2018jhn,Choudhury:2018byy,RoyChoudhury:2019hls,Vagnozzi:2018pwo,Vagnozzi:2019utt,Bonilla:2018nau} and  references therein). The standard parameters characterizing these effects are the effective number of neutrino species $N_{\rm eff}$ and the total neutrino mass scale $M_{\nu}$. We refer to \cite{Aghanim:2018eyx, DiValentino:2019dzu} for  the most recent constraints on these parameters. In principle, both the quantities $N_{\rm eff}$ and $M_{\nu}$, are model dependent, and hence, different cosmological scenarios may bound these parameters in different ways. In the context of a possible interaction between the dark components, the inclusion of neutrinos was investigated in \cite{Kumar:2016zpg,Feng:2019mym,Guo:2018gyo,Feng:2017usu,Guo:2017hea,Ghosh:2019tab}, where the presence of neutrinos can influence the coupling parameter between DE -- DM, and in the opposite direction, by assuming a dark coupling, the observational bound on the properties of neutrinos may minimally change  with respect to the minimal $\Lambda$CDM scenario. Moreover, beyond the three neutrino species of the standard model, the so called (3+1) neutrino model can also induce a non-null dark interaction \cite{Kumar:2017dnp}. 

How neutrinos can influence the free parameters of the models beyond the $\Lambda$CDM scenario, can open new perspectives to verify, in a more realistic way, the observational feasibility of the non-standard cosmological models. Such investigations are necessary in order to obtain more robust and accurate results on the full baseline of the alternative cosmological scenarios. In the present work, we consider an interacting vacuum energy scenario in presence of neutrinos and constrain it using the final Planck CMB observations, aiming to check how the inclusion of neutrinos may influence the coupling parameter and vice versa. Our results present a comprehensive updates on such interacting cosmological models in light of the CMB data from Planck's final release. 

The work has been organized in the following way. In section \ref{sec-2} we describe the basic equations for an interacting scenario in the background of a homogeneous and isotropic Universe.
Then in section \ref{sec-data} we describe the entire cosmological datasets and the statistical methodology to constrain the prescribed interacting scenarios in this work. The next section \ref{sec-results} is devoted to analyse the outcomes of the statistical results and their physical implications. 
Finally, in section \ref{sec-discuss} we close the present work offering a brief summary of the entire outcomes.

\section{Interaction in the dark sector}
\label{sec-2}

In the large scale, our Universe is almost homogeneous and isotropic and such geometrical configuration is characterized  by the Friedmann-Lema\^{i}tre-Robertson-Walker (FLRW) metric. We assume that the gravitational sector of our Universe  is described by the General theory of Relativity and the matter sector is minimally coupled to gravity. We further assume that in the matter sector, a non-gravitational interaction between the two main dark species of our Universe, namely, DM and DE exists. Here, DM is assumed to be pressure-less (i.e., cold) and DE is described by the vacuum energy density. The dark interaction is quantified as

\begin{eqnarray}
\label{DE_DM_1}
\nabla_{\mu}T_{i}^{\mu \nu }=Q_{i}^{\nu}\,, \quad \sum\limits_{\mathrm{i}}{%
Q_{i}^{\mu }}=0~,
\end{eqnarray}
where $i = c$ represents DM and $i=x$ represents DE. The four-vector $Q_{i}^{\mu}$ actually governs the interaction. We assume that $Q_{i}^{\mu}$ is given by 
\begin{eqnarray}
Q_{i}^{\mu}=(Q_{i}+\delta Q_{i})u^{\mu}+a^{-1}(0,\partial^{\mu}f_{i}), 
\end{eqnarray}
where $u^{\mu}$ is the velocity four-vector and $Q_i$ is the  background energy transfer. Let us note that from now on we shall use the notation $Q_i \equiv Q$.  The symbol $f_i$ refers to the momentum transfer potential. In the FLRW background, one can write down the conservation equations of the DM and DE densities as 

\begin{eqnarray}
&&\dot{\rho}_c +3 H \rho_c  = - Q~,\label{cont1}\\
&&\dot{\rho}_x = Q~,\label{cont2}
\end{eqnarray}
where $H  = \dot{a}/a$, is the rate expansion of the Universe. The symbol $H_0$ used hereafter in this work therefore refers to the present value of the Hubble parameter. 

In the present work, we consider a very well known (although phenomenological) parametric form of the interaction function $Q$, namely,

\begin{eqnarray}
\label{interaction-model}
Q = 3 H \xi \rho_x~,
\end{eqnarray}
where $\xi$ is the coupling parameter between the dark components. From the sign of $\xi$, one can identify the direction of the energy flow between the dark sectors. The condition $\xi < 0$ corresponds to the energy flow from DM to DE, and $\xi > 0$ represents the opposite scenario. The functional form $Q = 3 H \xi \rho_x$ can avoid the  instabilities in the perturbations at early times on the dark sector species. There are some other interaction models which could also remove the early time instabilities.  It should be noted that the inclusion of the global factor $H$ into the interaction function $Q$ was motivated to quantify a {\it possible global interaction} through the cosmic history. Although as already argued by the authors \cite{Valiviita:2008iv}, the interaction in the dark sectors should depend on the local quantities and not on the global expansion rate, namely, the Hubble factor. Nevertheless, as the present interaction function is widely studied in the literature before the final release of the Planck data, thus, we aim to revisit this model in order to see if the new Planck data sufficiently affect the key cosmological parameters within this interaction scenario, such as the coupling parameter of this interaction model and other parameters carrying the information about the neutrino properties of the Universe.  

On the other hand, the choice of the interaction function is not unique and it is very difficult to provide with a specific functional form since
the nature/properties of both dark components is completely unknown at present date. From the observational perspectives, we do not find any strong signal which could reveal the nature of the dark components, and hence, we believe that the nature of the interaction function will probably remain unknown for the next decade(s). We can only approximate the interaction function $Q$ only through the theoretical arguments and consistency with observational data. However, it has been argued by many investigators that the interaction between the dark components may appear from some effective field theory \cite{Gleyzes:2015pma,Boehmer:2015kta,Boehmer:2015sha}, disformal coupling \cite{vandeBruck:2015ida,Xiao:2018jyl},  axion monodromies \cite{Amico:2016qft}, varying
dark matter mass and fundamental constants \cite{Marsh:2017prd}, Horndeski theories \cite{Kase:2019hor}, or something else, such as the minimally varying Lambda theories \cite{Alexander:2019wne}. Thus, one can see that some viable action formalism can be given for the interacting dark energy theory. Motivated to check the strength of this dark coupling and its observational viability, let us assume in the present work the simple parametric function in eq. (\ref{interaction-model}) to quantify such dark interaction.

Our present methodology to describe the evolution of the linear perturbations of the dark component, $\delta_c$ and $\delta_x$, is the same as described in the earlier works, see for instance \cite{Wang:2014xca} (a thorough investigation in this direction can also be found in \cite{Guo:2017hea}). Here, we will do a brief review of the modified evolution of the dark species in the first order perturbation in presence of a non-gravitational interaction between them. In order to deal with the perturbations equations, we need the metric in its perturbed form. Let us consider the general perturbed FLRW metric is given by \cite{Mukhanov:1990me,Ma:1995ey,Malik:2008im}: 

\begin{eqnarray*}
\label{perturbed-metric}
ds^2 = -(1+ 2 \phi) dt^2 + 2 a \partial_i B dt dx^i + \\ 
a^2[(1-2\phi) \delta_{ij} + 2 \partial_i \partial_j E] dx^i dx^j,
\end{eqnarray*}
where $\phi$, $B$, $\psi$, $E$ are the gauge-dependent scalar perturbation quantities. We shall work in the synchronous gauge, that means, $\phi = B = 0$, $\psi = \eta$, and $k^2 E = -h/2 - 3 \eta$, with $k$ being the Fourier mode and $h$ and $\eta$ are the scalar metric perturbations. We refer to the works by \ref{Mukhanov:1990me,Ma:1995ey,Malik:2008im} for a precise description on the cosmological perturbations.

The components of interacting vacuum and DM, given in eq. (\ref{DE_DM_1}) reduce to the energy continuity equations

\begin{eqnarray}
\label{delta_DM}
\dot{\delta \rho_{c}} + 3 H \delta \rho_{c} - 3 \rho_c \dot{\psi} + \rho_{c} \frac{k^2}{a^2} \Big( \theta_{c} + a^2 \dot{E} \Big) = - \delta Q, 
\end{eqnarray}
\begin{eqnarray}
\label{delta_DE}
\dot{\delta \rho_{x}} = \delta Q,
\end{eqnarray}
and the momentum conservation equations lead to

\begin{eqnarray}
\label{theta_DM}
\rho_c \dot{\theta_{c}} = -f - Q(\theta - \theta_c),
\end{eqnarray}
\begin{eqnarray}
\label{theta_DE}
-\delta_{x} = f + Q \theta,
\end{eqnarray}
where $f = f_x = - f_c$ and $f$ is the momentum transfer. Thus, $f_x$, $f_c$ respectively stands for the momentum transfer associated to DE and DM. 
Now, one can combine eqns. (\ref{delta_DM}) and (\ref{delta_DE}), and eqns. (\ref{theta_DM}), (\ref{theta_DE}) in order to eliminate $\delta Q$ and $f$, and obtain

\begin{eqnarray}
\label{delta_DM_2}
\dot{\delta \rho_c} + 3 H \delta \rho_c - 3 \rho_c \dot{\psi} + \frac{k^2}{a^2} \Big(\theta_c + a^2 \dot{E} \Big) = - \dot{\delta_x}~,
\end{eqnarray}
\begin{eqnarray}
\label{theta_DM_2}
\rho_c \theta_c = \delta_x + Q \theta_c.
\end{eqnarray}
We consider an energy flow parallel to the 4-velocity of DM $Q^{\mu}_c = - Q^{\mu} u^{\mu}_c$. In this case, DM follows geodesics \cite{Wang:2014xca}. It means that the vacuum energy perturbations vanish in the DM-comoving frame from eq. (\ref{theta_DM_2}). Thus, from the residual gauge freedom in the synchronous gauge, we obtain $\theta_c = 0$ and $\delta_x = 0$. Therefore, in the comoving synchronous gauge, the density perturbation equation for DM then assume the final form

\begin{eqnarray}
\label{delta_DM_3}
\dot{\delta_c} = - \frac{h}{2} + \frac{Q}{\rho_c} \delta_c.
\end{eqnarray}

For all the non-interacting species, i.e., the photons, neutrinos and baryons, the linear perturbations evolution follows the standard evolution as predicted in the $\Lambda$CDM model. In what follows, we present the observational data sets as well as the methodology of the statistical analysis that we have adopted in this work.

\section{Data set and methodology}
\label{sec-data}

In this section, we briefly present the cosmological datasets and the statistical methodology to constrain all the interacting scenarios to be considered in this work. In what follows we describe the observational data.

\begin{enumerate}
    
\item Cosmic Microwave Background (CMB): We consider the latest cosmic microwave background measurements from Planck \cite{Aghanim:2018eyx,Aghanim:2018oex,Aghanim:2019ame}. The dataset is referred to  as Planck TT,TE,EE+lowE. Here we just refer this dataset as Planck 2018. 

\item Baryon acoustic oscillation (BAO) distance measurements: We have used distinct measurements of BAO data, such as the measurement from \footnote{In our analysis, we assume a sound horizon $r_{\rm d} = 147.21$ Mpc, as obtained by the Planck collaboration within the minimal $\Lambda$CDM cosmology. }:

(i) The 6dF Galaxy Survey, where BAO detection allows us to constrain the distance-redshift relation at $z_{\rm eff}=0.106$, with a spherically-averaged distance to $r_d/D_V(0.106) = 0.336 \pm 0.015$ Mpc \cite{Beutler:2011hx}.

(ii) The Main Galaxy Sample of Data Release 7 of the Sloan Digital Sky Survey, with BAO signal measured at $z_{\rm eff}=0.15$, with $D_V (z_{\rm eff}) = (664 \pm 25)(r_d/r_{\rm d,fid})$ Mpc ~\cite{Ross:2014qpa}. In this BAO measure is assumed $r_{\rm d,fid}= 148.69$ Mpc \cite{Ross:2014qpa}.

(iii) The BAO Spectroscopic Survey data release (DR12), to the distance scale:
$D_V (0.38) = (1477 \pm 16) (r_d/r_{\rm d,fid})$ Mpc, $D_V (0.51) = (1877 \pm 19 ) (r_d/r_{\rm d,fid})$ Mpc and $D_V (0.61) = (2140 \pm 22) (r_d/r_{\rm d,fid})$ Mpc, \cite{Alam:2016hwk}. In obtaining these measures are assumed $r_{\rm d,fid} = 147.78$ Mpc \cite{Alam:2016hwk}.

The distance scale $D_V$ in the above measures is defined as
\begin{eqnarray}
D_V(z) = \Big( D^2_M(z) \frac{cz}{H(z)} \Big)^{1/3},
\end{eqnarray}
where $D_M$ is the comoving angular diameter distance.

These BAO modeling are geometrical quantity. Thus, the dependence on the cosmological model in our statistical analysis will be only on the $H(z)$ function (directly and intrinsically via $D_M$ integration), which here is modified by the presence of a dark interaction. Therefore, both the coupling parameter $\xi$, as well as the neutrino properties in terms of $M_{\nu}$ and $N_{\rm eff}$, are bound by its dependence on the $H(z)$ function only. These BAO points are essentially constructed assuming $\Lambda$CDM as the fiducial model. We did not modify any other cosmology and/or we did not consider the modeling of nonlinear scales effects in order to obtain the data listed above. 

\item Hubble Space Telescope (HST): We include the very latest measurement of the Hubble constant from the Hubble Space Telescope, yielding $H_0 = 74.03 \pm 1.42$ km/s/Mpc at $68\%$ CL~\cite{Riess:2019cxk}. This estimation of the Hubble constant  is in tension ($4.4 \sigma$) with Planck's estimation within the minimal $\Lambda$CDM model. In this work we refer to this data  as R19.  

\end{enumerate}

Concerning the statistical analyses, we modify the Markov Chain 
Monte Carlo code  \texttt{cosmomc} \cite{Lewis:2002ah,Lewis:1999bs}, in order to extract the observational constraints for the interaction model described in the previous section. In this work, we consider three different scenarios as follows:
\bigskip 

\textbf{Scenario 1}: We consider a minimal scenario of dark sectors' interaction, being characterized only with a single parameter, the coupling parameter $\xi$, beyond the $\Lambda$CDM model baseline. The parameter space of this model is,

\begin{eqnarray}
\mathcal{P} \equiv\Bigl\{\Omega_{b}h^2, \Omega_{c}h^2, 100\theta_{MC}, \tau, n_{s}, log[10^{10}A_{s}], \xi \Bigr\}~,
\label{eq:parameter_space1}
\end{eqnarray}
which is seven dimensional. We label this model as Interacting Vacuum Scenario (IVS).
\bigskip 

\textbf{Scenario 2}: We consider an extended parameter space by including the neutrino mass scale $M_{\nu}$  as a free parameter. Thus, the model baseline is given by

\begin{eqnarray}
\mathcal{P} \equiv\Bigl\{\Omega_{b}h^2, \Omega_{c}h^2, 100\theta_{MC}, \tau, n_{s}, \log[10^{10}A_{s}], 
\xi, M_{\nu}\Bigr\},
\label{eq:parameter_space2}
\end{eqnarray}
which is eight dimensional. We label this scenario as IVS + $M_{\nu}$ following the earlier labeling. 

Regarding the sum of neutrino masses, we impose a prior of $M_{\nu} > 0$, ignoring a possible lower limit from neutrino oscillation experiment and assuming three neutrinos species, that is, $N_{\rm eff} = 3.046$. For the purposes of obtaining bounds on neutrino mass from the cosmological data, the prior $M_{\nu} > 0$ is adequate. 
\bigskip 

\textbf{Scenario 3}: As a third and final scenario in this work, we consider the extended parameter space including both $M_{\nu}$ and $N_{\rm eff}$ as free parameters. Thus, in this generalized case the parameter space of our interest is,

\begin{eqnarray}
\mathcal{P} \equiv\Bigl\{\Omega_{b}h^2, \Omega_{c}h^2, 100\theta_{MC}, \tau, n_{s}, \log[10^{10}A_{s}], \nonumber\\ 
\xi, M_{\nu}, N_{\rm eff} \Bigr\},
\label{eq:parameter_space3}
\end{eqnarray}
which is nine dimensional and we label it as IVS + $M_{\nu}$ $+$ $N_{\rm eff}$. 

We consider the flat FLRW Universe in this work, which is clear from the above parameter spaces. During the statistical analyses, we consider the flat priors on all parameters,
the common baseline parameters in all scenarios is taken to be: $\Omega_{b} h^2 \in [0.005\,,\,0.1]$, $\Omega_{c} h^2 \in [0.01\,,\,0.99]$, $\theta_{{MC}} \in [0.5\,,\,10]$, $\tau \in [0.01\,,\,0.8]$, $\log_{10}(10^{10}A_{s}) \in [2\,,\,4]$ and $n_s \in [0.8\,,\, 1.2]$. For all interacting scenarios, we take $\xi \in [-1 \,,\, 1]$. For the scenario IVS + $M_{\nu}$, we impose $M_{\nu} \in [0, 1]$. For the model IVS + $M_{\nu}$ $+$ $N_{\rm eff}$, we assume $M_{\nu} \in [0, 1]$ and $N_{\rm eff} \in [1, 10]$. Note that the prior on the total neutrino mass $M_{\nu}$ is compatible with its most recent bounds from the Karlsruhe Tritium Neutrino Experiment (KATRIN) measures \cite{KATRIN}. 
Let us now discuss the main observational results extracted using the above observational data. 

\begingroup                                                                                                                     
\squeezetable                                                                                                                   
\begin{center}                                                                                                                  
\begin{table*}                                                                                                                   
\begin{tabular}{ccccccccc}                                                                                                            
\hline\hline                                                                                                                    
Parameters & Planck 2018 & Planck 2018 + BAO & Planck 2018 + R19 \\ \hline

$\Omega_c h^2$ & $    0.0687_{-    0.0677-    0.0677}^{+    0.0244+    0.0647}$ & $    0.0996_{-    0.0156-    0.0383}^{+    0.0225+    0.0353}$ & $    0.0378_{-    0.0346-    0.0368}^{+    0.0118+    0.0476}$ \\

$\Omega_b h^2$ & $    0.02230_{-    0.00015-    0.00029}^{+    0.00015+    0.00030}$ & $    0.02233_{-    0.00014-    0.00027}^{+    0.00014+    0.00028}$ & $    0.02233_{-    0.00015-    0.00028}^{+    0.00014+    0.00031}$ \\

$100\theta_{MC}$ & $    1.04409_{-    0.00405-    0.00493}^{+    0.00258+    0.00548}$ & $    1.04188_{-    0.00134-    0.00207}^{+    0.00086+    0.00233}$ & $    1.04625_{-    0.00180-    0.00381}^{+    0.00233+    0.00357}$ \\

$\tau$ & $    0.054_{-    0.0079-    0.015}^{+    0.0075+    0.015}$ & $    0.055_{-    0.0083-    0.016}^{+    0.0076+    0.016}$ & $    0.054_{-    0.0080-    0.016}^{+    0.0081+    0.016}$ \\

$n_s$ & $    0.9723_{-    0.0044-    0.0081}^{+    0.0043+    0.0083}$ & $    0.9734_{-    0.0040-    0.0078}^{+    0.0040+    0.0079}$  & $    0.9734_{-    0.0043-    0.0082}^{+    0.0043+    0.0083}$ \\

${\rm{ln}}(10^{10} A_s)$ & $    3.055_{-    0.016-    0.030}^{+    0.015+    0.031}$ & $    3.057_{-    0.017-    0.032}^{+    0.016+    0.033}$ & $    3.054_{-    0.016-    0.034}^{+    0.016+    0.033}$ \\

$\xi$ & $    0.132_{-    0.077-    0.197}^{+    0.142+    0.169}$ & $    0.059_{-    0.061-    0.101}^{+    0.053+    0.110}$ & $    0.207_{-    0.034-    0.132}^{+    0.067+    0.092}$ \\

$\Omega_{m0}$ & $    0.191_{-    0.141-    0.166}^{+    0.075+    0.191}$ & $    0.261_{-    0.046-    0.099}^{+    0.056+    0.095}$  & $    0.115_{-    0.065-    0.085}^{+    0.031+    0.118}$ \\

$H_0$ & $   70.84_{-    2.50-    5.94}^{+    4.26+    5.26}$ & $   68.82_{-    1.53-    2.64}^{+    1.30+    2.77}$ & $   73.27_{-    1.08-    3.13}^{+    1.87+    2.65}$ \\

\hline\hline                                                                                                                    
\end{tabular}                                                                                                                   
\caption{Observational constraints at 68\% and 95\% CL on free and derived parameters of the IVS scenario.  The parameter $H_0$ is in the units of km/sec/Mpc. } 
\label{tab:IVS1}                                               
\end{table*}                                                    
\end{center}                                                   
\endgroup  
 \begingroup                                                                                                                     
\squeezetable                                                                                                                   
\begin{center}                                                                                                                  
\begin{table*}                                                                                                                   
\begin{tabular}{ccccccccccccccccc}                                                                                                            
\hline\hline                                                                                                                    
Parameters & Planck 2018 & Planck 2018 + BAO & Planck 2018 + R19\\ \hline

$\Omega_c h^2$ & $    0.0816_{-    0.0371-    0.0806}^{+    0.0569+    0.0621}$ & $     0.1012_{-    0.0173-    0.0427}^{+    0.0248+    0.0395}$ & $    0.033_{-    0.0249-    0.0318}^{+    0.0145+    0.0334}$ \\

$\Omega_b h^2$ & $    0.02227_{-    0.00016-    0.00032}^{+    0.00015+    0.00030}$ & $    0.02233_{-    0.00014-    0.00029}^{+    0.00014+    0.00028}$ & $    0.02231_{-    0.00015-    0.00029}^{+    0.00015+    0.00030}$ \\

$100\theta_{MC}$ & $    1.04315_{-    0.00392-    0.00476}^{+    0.00199+    0.00576}$ & $    1.04181_{-    0.00147-    0.00233}^{+    0.00096+    0.00263}$  & $    1.04661_{-    0.00165-    0.00302}^{+    0.00175+    0.00312}$  \\

$\tau$ & $    0.055_{-    0.0085-    0.015}^{+    0.0074+    0.016}$ & $    0.056_{-    0.0082-    0.015}^{+    0.0076+    0.016}$ & $    0.055_{-    0.0079-    0.016}^{+    0.0078+    0.016}$  \\

$n_s$ & $    0.9711_{-    0.0047-    0.0091}^{+    0.0044+    0.0094}$ & $    0.9728_{-    0.0041-    0.0081}^{+    0.0043+    0.0084}$  & $    0.9722_{-    0.0042-    0.0083}^{+    0.0043+    0.0083}$  \\

${\rm{ln}}(10^{10} A_s)$ & $    3.057_{-    0.017-    0.030}^{+    0.016+    0.033}$ & $    3.057_{-    0.016-    0.031}^{+    0.015+    0.033}$  & $    3.055_{-    0.016-    0.033}^{+    0.016+    0.033}$  \\

$\xi $ & $    0.102_{-    0.097-    0.202}^{+    0.152+    0.180}$ & $    0.054_{-    0.069-    0.114}^{+    0.059+    0.126}$  & $    0.220_{-    0.037-    0.089}^{+    0.053+    0.082}$  \\

$\Omega_{m0}$ & $    0.227_{-    0.147-    0.174}^{+    0.095+    0.191}$ & $    0.265_{-    0.052-    0.109}^{+    0.060+    0.100}$  & $    0.104_{-    0.049-    0.067}^{+    0.028+    0.075}$  \\

$H_0$ & $   69.45_{-    3.47-    6.07}^{+    4.02+    5.73}$ & $   68.71_{-    1.45-    2.74}^{+    1.43+    2.79}$  & $   73.46_{-    1.12-    2.32}^{+    1.17+    2.28}$  \\

$M_\nu$ & $    < 0.277$ & $ < 0.158$  & $   < 0.204 $  \\

\hline\hline                                                                                                                    
\end{tabular}                                                                                                                   
\caption{Constraints at 68\% and 95\% CL on free and derived parameters of the IVS $+$ $M_{\nu}$ scenario. The parameter $H_0$ is measured in the units of km/s/Mpc, whereas $M_\nu$, reported in the 95\% CL, is in the units of eV. }
\label{tab:IVS2}                                                                                                   
\end{table*}                                                                                                                     
\end{center}                                                                                                                    
\endgroup                                                                                                                       
\begingroup                                                                                                                     
\squeezetable                                                                                                                   
\begin{center}                                                                                                                  
\begin{table*}                                                                                                                   
\begin{tabular}{ccccccccccc}                                                                                                            
\hline\hline                                                                                                                    
Parameters & Planck 2018 & Planck 2018 + BAO & Planck 2018 + R19 \\ \hline

$\Omega_c h^2$ & $    0.0758_{-    0.0748-    0.0748}^{+    0.0259+    0.0642}$ & $    0.1029_{-    0.0190-    0.0465}^{+    0.0265+    0.0446}$ & $    0.0346_{-    0.0336-    0.0336}^{+    0.0089+    0.0402}$   \\

$\Omega_b h^2$ & $    0.02221_{-    0.00023-    0.00045}^{+    0.00023+    0.00044}$  & $    0.02232_{-    0.00021-    0.00042}^{+    0.00021+    0.00041}$  & $    0.02232_{-    0.00019-    0.00037}^{+    0.00020+    0.00037}$  \\

$100\theta_{MC}$ & $    1.04362_{-    0.00414-    0.00497}^{+    0.00228+    0.00590}$  & $    1.04173_{-    0.00167-    0.00277}^{+    0.00108+    0.00302}$  & $    1.04647_{-    0.00168-    0.00370}^{+    0.00244+    0.00351}$  \\

$\tau$ & $    0.055_{-    0.0072-    0.015}^{+    0.0072+    0.015}$  & $    0.055_{-    0.0084-    0.015}^{+    0.0077+    0.017}$ & $0.055_{-    0.0075-    0.015}^{+    0.0076+    0.016}$  \\

$n_s$ & $    0.9686_{-    0.0088-    0.0172}^{+    0.0085+    0.0166}$   & $    0.9727_{-    0.0081-    0.0156}^{+    0.0080+    0.0160}$ & $    0.9733_{-    0.0071-    0.0136}^{+    0.0071+    0.0140}$  \\

${\rm{ln}}(10^{10} A_s)$ & $    3.053_{-    0.018-    0.036}^{+    0.018+    0.036}$  & $    3.056_{-    0.020-    0.035}^{+    0.018+    0.038}$  & $    3.057_{-    0.017-    0.034}^{+    0.017+    0.034}$ \\

$\xi$ & $    0.116_{-    0.093-    0.213}^{+    0.155+    0.189}$ & $    0.049_{-    0.074-    0.132}^{+    0.061+    0.131}$  & $  0.217_{-    0.035-    0.101}^{+    0.064+    0.087}$ \\

$\Omega_{m0}$ & $    0.215_{-    0.153-    0.184}^{+    0.091+    0.202}$ & $    0.269_{-   0.052-    0.113}^{+    0.063+    0.114}$ & $    0.108_{- 0.058-  0.072}^{+   0.027+    0.086}$ \\

$H_0$ & $   69.46_{-    3.77-    6.37}^{+    4.34+    6.29}$   & $   68.58_{-    1.59-    3.14}^{+    1.60+    3.22}$ &  $   73.50_{-    1.22-    2.44}^{+    1.35+    2.41}$ \\

$M_\nu$ & $    < 0.272$  & $ < 0.164$ & $ < 0.201 $  \\

$N_{\rm eff}$ & $    2.98_{-    0.20-    0.36}^{+    0.19+    0.37}$ & $    3.05_{-    0.20-    0.37}^{+    0.18+    0.37}$ & $    3.08_{-    0.17-    0.32}^{+    0.17+    0.34}$  \\

\hline\hline                                                
\end{tabular}                                                   
\caption{Constraints at 68\% and 95\% CL on free and derived parameters of the IVS $+$  $M_{\nu}$ $+$ $N_{\rm eff}$ scenario. The parameter $H_0$ is measured in the units of km/s/Mpc, whereas $M_\nu$ reported in the 95\% CL, is in the units of eV. }
\label{tab:IVS3}                                               
\end{table*}                                                    
\end{center}                                                                                                                    
\endgroup                                                                                                                       

\begingroup                                                                                                                     
\squeezetable

\endgroup

\section{Results}
\label{sec-results}

In this section, we report and discuss the observational constraints on the dark coupling parameter $\xi$, as well as the full parameter space of the model, assuming three different interacting scenarios described in the previous section \ref{sec-data}, using CMB measurements from Planck 2018, BAO data and the local estimation of the Hubble constant, $H_0$, from HST. Thus, one  can easily check how these three data sets, in particular, the CMB data from Planck 2018 release may bound a possible DM--DE coupling, in its minimal parametric space and also in its possible extensions including the neutrinos.

Tables \ref{tab:IVS1}, \ref{tab:IVS2} and \ref{tab:IVS3} summarize the main results of the statistical analyses carried out using three different datasets, such as, Planck 2018, Planck 2018 + BAO, and Planck 2018 + R19 for three distinct interacting scenarios, namely, IVS, IVS $+$ $M_{\nu}$ and IVS $+$ $M_{\nu}$ $+$ $N_{\rm eff}$, respectively. We also consider the joint analysis Planck 2018 + R19, because as we will see, when analyzing with Planck 2018 data only, the scenario under consideration here has no tension with local $H_0$ measurements by HST. The inclusion of BAO data is motivated to break the degeneracy on the full parametric space of the scenarios when analyzing with Planck 2018 data only. Differences between these combinations will be discussed in this section.
\begin{figure}
\centering
\includegraphics[width=0.43\textwidth]{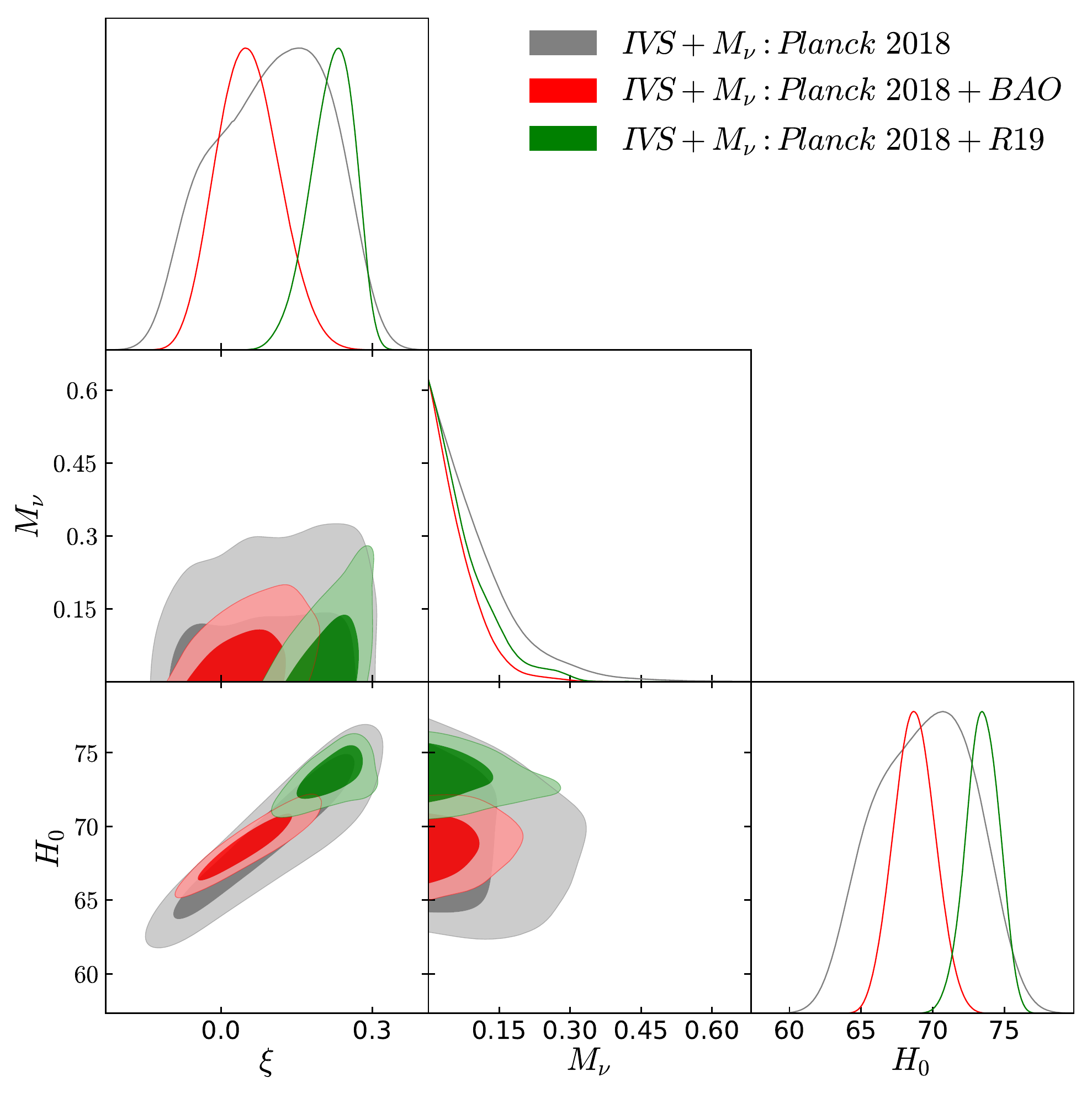}
\caption{The 1D marginalized posterior distributions and the 2D parametric space at 68\% CL and 95\% CL for the scenario IVS + $M_{\nu}$. }
\label{PS_IVS_Mnu}  
\end{figure}
\begin{figure}
\centering
\includegraphics[width=0.43\textwidth]{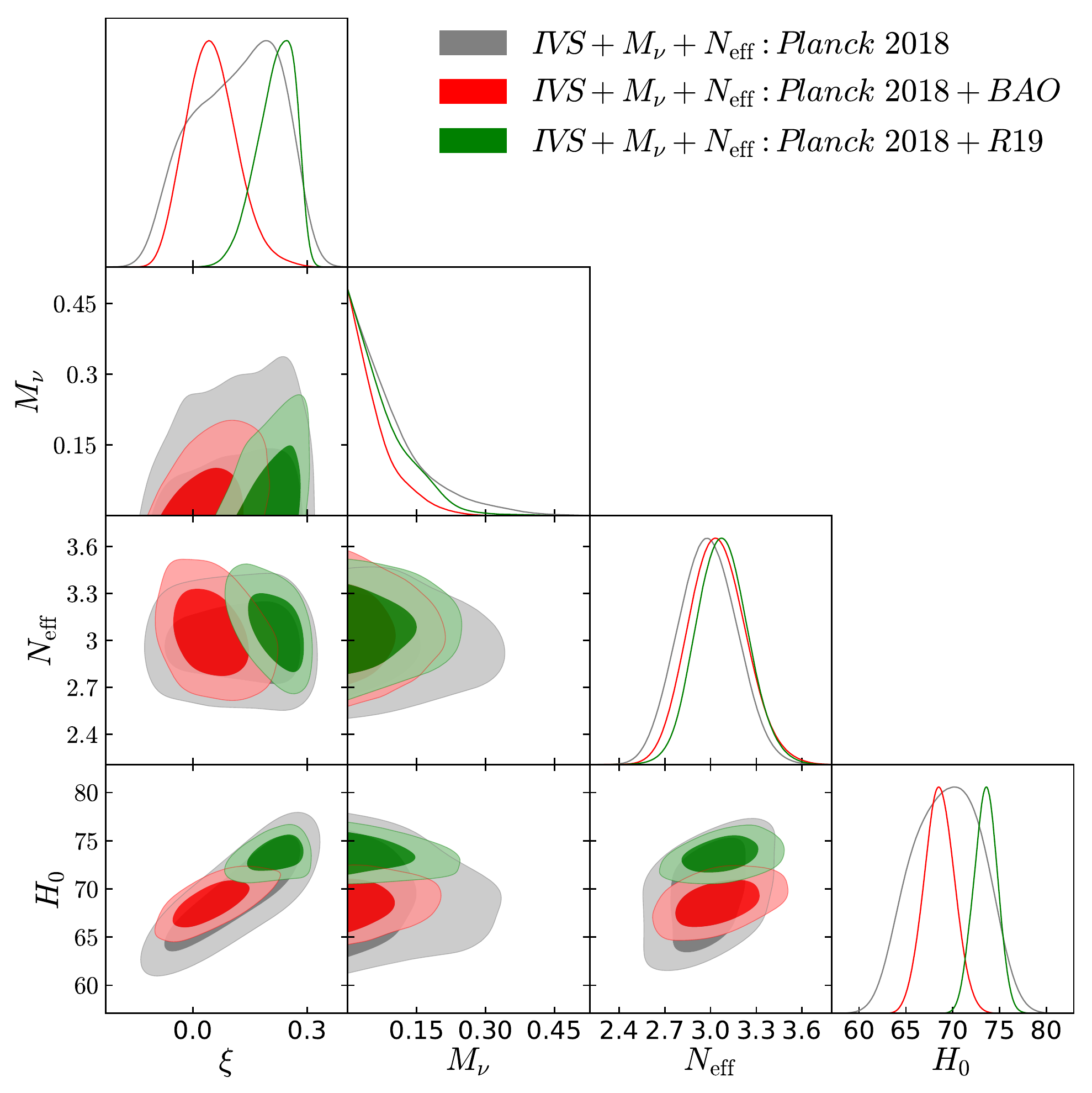}
\caption{Same as figure \ref{PS_IVS_Mnu}, but assuming the IVS + $M_{\nu} + N_{\rm eff}$ scenario. }
\label{PS_IVS_Mnu_Neff}  
\end{figure}
\begin{figure*}
\centering
\includegraphics[width=0.32\textwidth]{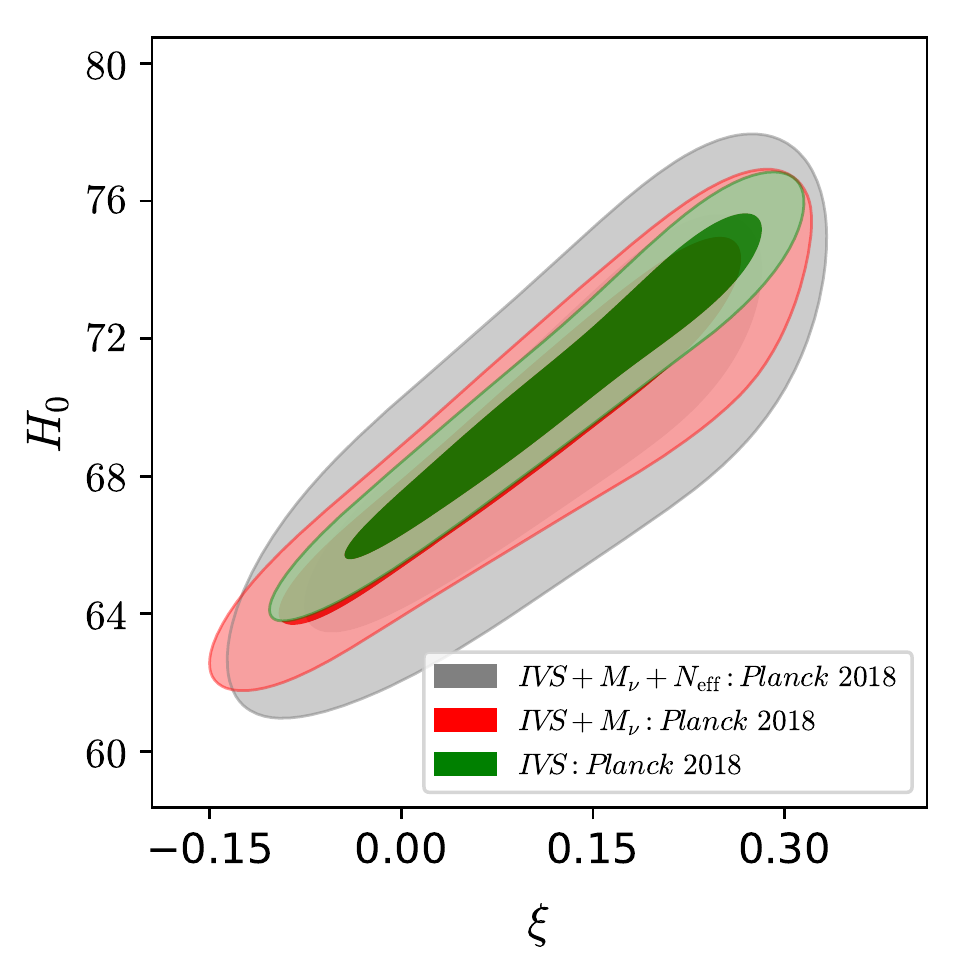}
\includegraphics[width=0.32\textwidth]{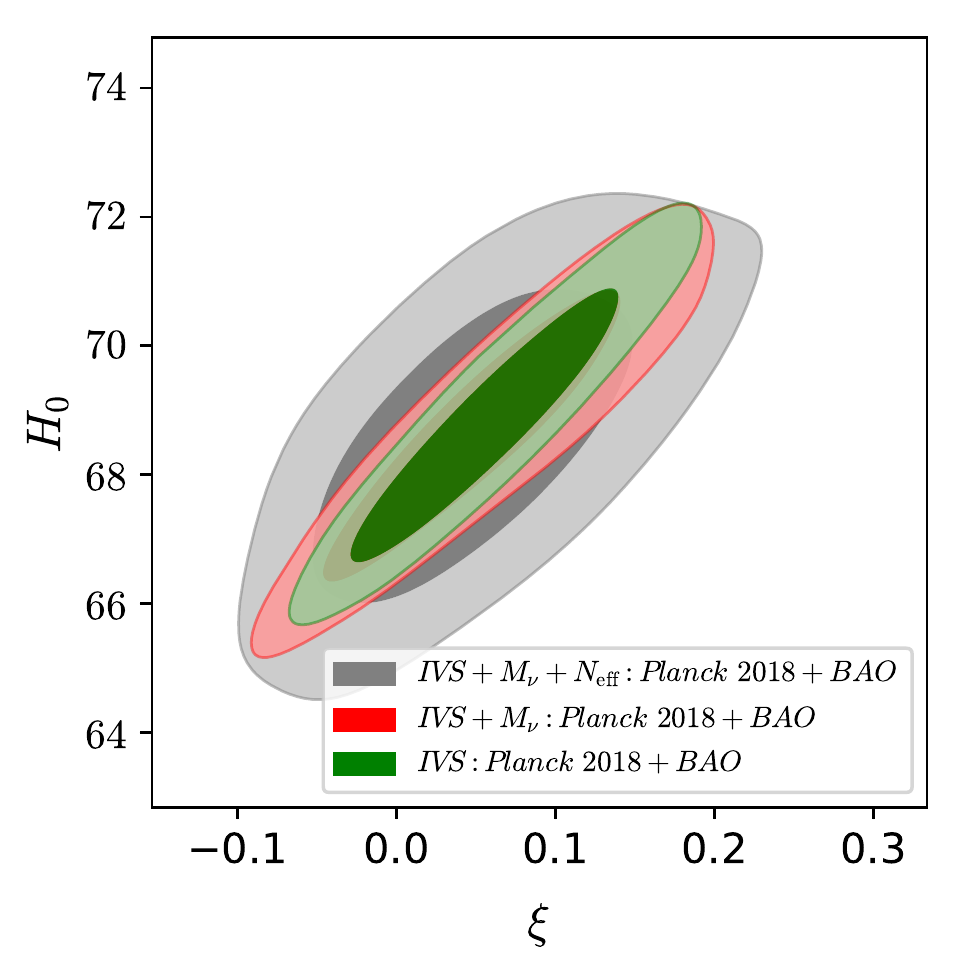}
\includegraphics[width=0.32\textwidth]{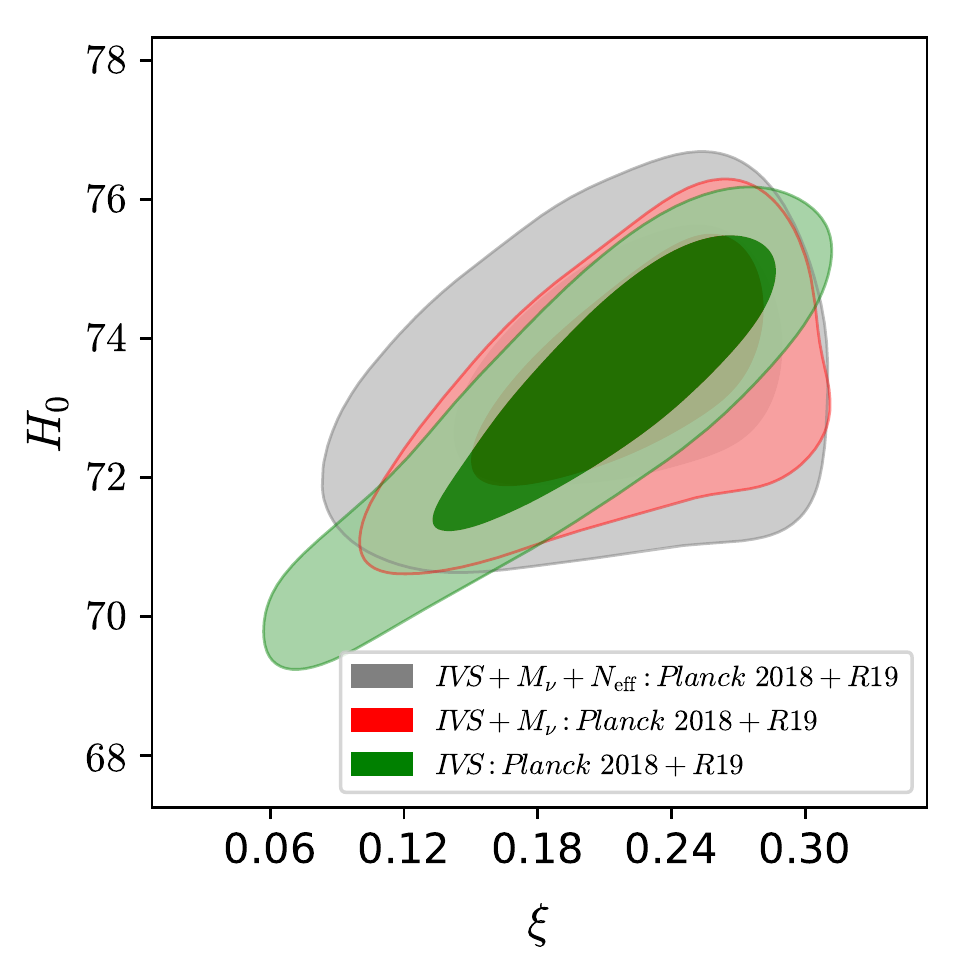}
\caption{Parametric space in the plane $\xi$ - $H_0$ comparing different scenarios under the same data set. Left panel: Planck 2018 only. Middle panel: Planck 2018 + BAO. Right panel: Planck 2018 + R19. }
\label{2d_H0_xi}  
\end{figure*}
the Figs. \ref{PS_IVS_Mnu} and \ref{PS_IVS_Mnu_Neff}, respectively show the parametric space at 68\% CL and 95\% CL for some selected parameters of the interacting scenarios, IVS + $M_{\nu}$ and IVS + $M_{\nu}$ + $N_{\rm eff}$, for the mentioned datasets. Under this specific dark sector interaction, we note that the inclusion of $M_{\nu}$ and $N_{\rm eff}$, does not affect the observational perspectives of the coupling parameter $\xi$ {\bf (in the sense to have a potential indication for $\xi \neq 0$)}. One can observe that the coupling parameter $\xi$ is very weakly correlated with $M_{\nu}$ and $N_{\rm eff}$, and the differences on the statistical confidence of $\xi$ are essentially due to the differences induced for the data combinations. Therefore, the presence of $M_{\nu}$ and $N_{\rm eff}$ does not significantly shift  the prediction on the dark coupling parameter. On the other hand, while considering the possibility of such dark coupling, within the IVS + $M_{\nu}$ model, we find that $M_{\nu} < 0.27$ eV for CMB data only, while $M_{\nu} < 0.12$ eV within the minimal $\Lambda$CDM model \cite{Aghanim:2018eyx}. The inclusion of BAO and R19 data to Planck 2018, offers more tight constraint on the total neutrino mass as, $M_{\nu} < 0.164$ eV (95\% CL, Planck 2018 + BAO) and $M_{\nu} < 0.201$ eV (95\% CL, Planck 2018 + R19). The bounds on $M_{\nu}$ within IVS + $M_{\nu}$ + $N_{\rm eff}$ scenario remains practically unchanged compared to IVS + $M_{\nu}$ case. {\bf We find that} $N_{\rm eff} = 2.98^{+0.37}_{-0.36}$ (at 95\% CL) within the IVS + $M_{\nu}$ + $N_{\rm eff}$ scenario from CMB data, while $N_{\rm eff} = 2.92^{+0.36}_{-0.37}$ (95\% CL) within the minimal $\Lambda$CDM \cite{Aghanim:2018eyx}. Thus, the presence of such dark coupling predicts a significant difference on the neutrino mass scale, but does not affect the effective number of neutrino species. Similar conclusions are drawn from other data combinations considered here. Once that $M_{\nu}$ -- $\xi$, is very weakly correlated, the effects of a possible dark coupling can be quantified as a possible increase on the neutrino mass scale predicting three active neutrinos. The reverse way, that is, the presence of massive neutrinos does not change the perspectives on the coupling parameter under the model in consideration.

Recently, the KATRIN experiment \cite{KATRIN} reports that the total neutrino mass is not larger than  $1.1$ eV  (at 90\% CL), that means, $M_{\nu} < 1$ eV. 
But, as we can see, the upper bounds on the total neutrino mass from the cosmological information are still the most restrictive, putting $M_{\nu} < 0.12$ eV within $\Lambda$CDM model and $M_{\nu} < 0.164$ eV from the IVS scenarios, both from Planck 2018 + BAO data combination at 95\% CL, for example. Therefore, we can see that the cosmological bounds on $M_{\nu}$ are fully compatible with the limits obtained from terrestrial neutrino experiments. 

In appendix \ref{appendix}, we show the observational constraints on a non-interacting cosmological model $w$CDM model with and without neutrinos,
using the same data combinations, motivated to check if the results can be mimicked by a very different assumption/model extension that is not the interacting DE scenario. We find that within this non-interacting $w$CDM + $M_{\nu}$  scenario,  the bounds on the total neutrino mass are as follows: $M_{\nu} < 0.287 \,\; (< 0.184) \,\; (< 0.296)$ eV at 95\% CL for CMB  (Planck 2018 + BAO), (Planck 2018 + R19), respectively. We notice that the bound obtained from the CMB data only, in direct comparison with IVS + $M_{\nu}$ scenario, is very compatible and both the interacting and non-interacting  models provide with the same limit on the neutrino mass scale. On the other hand, in view of both the joint analyses, i.e., Planck 2018 + BAO and Planck 2018 + R19, the bounds on $M_{\nu}$ are significantly wider compared to the $w$CDM model. Thus, the above observation shows that $M_{\nu}$ scale can be minimally model-dependent. 

Now, considering a more wider picture $w$CDM + $M_{\nu}$ + $N_{\rm eff}$, we find that $M_{\nu} < 0.314 \,\; (< 0.193) \,\; (< 0.324)$ eV at 95\% CL from CMB  (Planck 2018 + BAO), (Planck 2018 + R19), respectively. Compared to the $w$CDM + $M_{\nu}$ scenario, we see that in $w$CDM + $M_{\nu}$ + $N_{\rm eff}$ scenario, the bound on $M_{\nu}$ is minimally extended and this apparent difference seems to be a statistical fluctuation due to the MCMC analysis. On the other hand, when the scenario $w$CDM + $M_{\nu}$ + $N_{\rm eff}$ is compared to the IVS + $M_{\nu}$ + $N_{\rm eff}$ scenario, the non-interacting scenario $w$CDM predicts a bigger bound on the total neutrino mass. Thus, from the direct comparison between these models, i.e., $w$CDM and IVS cosmology, one can conclude that the bounds on $M_{\nu}$ can be model-dependent and hence different scenarios return different constraints on $M_{\nu}$. Concerning the $N_{\rm eff}$ parameter within this model scenario (i.e., $w$CDM + $M_{\nu}$ + $N_{\rm eff}$),  we notice that significant differences appear between the best fit values only for the joint analysis Planck 2018 + R19. However, when we consider present the interacting and the non-interacting scenarios, the constraints on $N_{\rm eff}$ are found to be compatible  within 68\% CL. So, no tension on the estimations of $N_{\rm eff}$ from these cosmological scenarios is observed.

Our updated constraints show that for all three interacting DE scenarios, the final Planck 2018 data can successfully solve the $H_0$ tension/problem. This dark coupling between DE and DM can generate high values of the Hubble constant $H_0$, in contrast to the $\Lambda$CDM case, where  it is not possible to obtain such high value of $H_0$. This aspect has previously been noted in such scenarios \cite{Kumar:2016zpg,Kumar:2017dnp,DiValentino:2017iww,Yang:2018euj,Yang:2018uae,Kumar:2019wfs,Pan:2019gop,DiValentino:2019ffd}. Under any general modification from $\Lambda$CDM cosmology, it is expected that the main effects on CMB anisotropies happen on the amplitude of the late time integrated Sachs-Wolfe effect (manifested at large angular scales), which depends on the duration of the dark energy-dominated stage, i.e., on the time of equality between matter and DE density, fixed by the ratio $\Omega_{x}/\Omega_{m}$, where $\Omega_{m} = \Omega_{c} + \Omega_{b}$. A larger $\Omega_{x}$ implies a longer DE domination, and consequently, an enhanced  late time integrated Sachs-Wolfe effect. It is important to note that in DE--DM coupling models, depending on different coupling functions $Q$,  different constraints on $\Omega_{c}$ can be achieved. The constraints on baryon density should remain practically unchanged. In our models under consideration, we are assuming our Universe to be spatially flat. Thus, $\Omega_{x} \simeq 1 - \Omega_{c}$, at late times. And the changes induced by different constraints on $\Omega_{c}$ will control mainly this correction on the CMB anisotropies at large scales. On the other side, $\Omega_{c}$ will control the amplitude of the third peak in the CMB spectra and also the constraints on $H_0$ in the form of $h = \sqrt{w_m (1 - \Omega_{x}})$, where $h = H_0/100$ and $w_m = h^2(\Omega_{b} + \Omega_{c})$. Moreover, the changes in the expansion of the Universe, induced mainly for the parameters $\xi$ and $H_0$ will contribute to the corrections on the amplitude of all peaks  and shifts the CMB spectrum due to the modifications in the angular diameter distance at decoupling (this effect depends on the expansion history due to the DE--DM interaction after the decoupling). The magnitude of the correlations are proportional to the possible deviations from $\xi = 0$, compared to the non-interacting $\Lambda$CDM model. From the results summarized in Table \ref{tab:IVS1}, \ref{tab:IVS2}, \ref{tab:IVS3}, it is notable that a possible presence of $\xi$ will generate these effects by widening the parameter space in order to obtain high $H_0$ values, especially looking at CMB constraints only. The presence of $M_{\nu}$ and $N_{\rm eff}$, has less effects on CMB than the presence of a possible coupling parameter $\xi$. In fact, these parameters (which characterize the  properties of neutrinos) do not shift the best fit values of $\xi$ significantly. Thus, the contribution to high $H_0$ values comes from the dark coupling parameter $\xi$. Once our global constraint on $H_0$ is not in tension with local $H_0$ measurement from HST data, we can safely consider the combination Planck 2018 + R19.  From the combined analysis Planck 2018 + R19, we find that the coupling parameter between DM--DE may be non-null at high statistical significance, i.e., at $> 3\sigma$, from Planck 2018 + R19. The natural explanation for having such non-null value of $\xi$ in high statistical significance is simple. Once the parametric space $H_0$ -- $\xi$ from Planck 2018 data alone is large (and compatible with high $H_0$ values) and strongly correlated, introducing the HST prior in the analysis, having high $H_0$ value with high accuracy, it will eliminate statistically a significant portion of the parameter space of $\xi$ preferring non-null values on this parameter. In general lines, we can interpret this solution for the $H_0$ problem by including the R19 prior in the joint analysis, leading to a new physics favoring a dark coupling. When analyzing with CMB from Planck 2018 only, we do not find any evidence for the coupling parameter, and $\xi = 0$ is found to be fully compatible.

We also analyze the combination Planck 2018 + BAO, in order to break the statistical degeneracy in the space of parameters, especially on $H_0$ and $\xi$. We find that the inclusion of the BAO data changes the paradigm compared to the joint analysis Planck 2018 + R19. Both the analyses (i.e., Planck 2018 + BAO and Planck 2018 + R19) are incompatible within 1$\sigma$, but fully compatible within 2$\sigma$. The analysis with Planck 2018 + BAO does not lead to any evidence for $\xi \neq 0$ and also does not assuage the high $H_0$ values from HST observations at less than 2$\sigma$. The combination Planck2018 + BAO and Planck 2018 + R19, bias the analysis with CMB only in opposite directions when compared to the mean values obtained from Planck 2018 data only (see figures \ref{PS_IVS_Mnu} and \ref{PS_IVS_Mnu_Neff}). It makes clear that different data combinations may return different perceptions on the coupling parameter, especially when various observational data are considered in the joint analysis. Since our main goal is to update the constraints under the perspective of the final Planck CMB data release, thus, let us just stick to the joint analysis Planck 2018 + BAO and Planck2018 + R19, within the motivation of these scenarios.

As mentioned earlier, the condition for a possible DE--DM coupling leads to a degeneracy and strong correlation in the $\xi$ -- $H_0$ plane. To make it easier for readers to understand, we quantify it in Fig. \ref{2d_H0_xi}, where one may note how different scenarios respond to the same data set. Also, quantifying some previous comments, one can argue that the presence of neutrinos does not significantly shift the mean value of the coupling parameter. In the right graph of Fig. \ref{2d_H0_xi}, $\xi > 0$ is clearly evident with high significance.

\section{Final remarks}
\label{sec-discuss}

In this article we have updated the observational constraints in light of the CMB data from final Planck release on cosmological scenarios motivated by a possible dark interaction between DE--DM, where DE is described by the vacuum energy density and DM is pressure-less. As a new ingredient in the analysis, we consider the neutrino properties within this interacting scenario. Concerning the main upgraded results, we found that while analyzing with CMB data from  Planck 2018, the constraint on $H_0$ is compatible with the estimation obtained from HST. For the joint analysis, Planck 2018 + R19, we note that the coupling parameter, $\xi$, is non-null at more than 3$\sigma$ CL. Results are strongly linked with the $H_0$ prior. Concerning the analyses with Planck2018 only and Planck 2018 + BAO, we do not find any strong evidence for a possible dark coupling in the Universe sector. We also notice that the inclusion of neutrino properties in terms of the parameters, namely, $M_{\nu}$ and $N_{\rm eff}$, does not significantly correlate with the coupling parameter $\xi$. Therefore, such neutrino properties do not directly influence the dark sectors' properties, at least within the present framework.

As discussed in other works of the literature, the presence of a non-gravitational coupling between DE and DM can solve the $H_0$ tension reported by Planck within the minimal $\Lambda$CDM model, as well as explain the observable Universe on all scales. Additionally, the initial theoretical motivations for such models together with the current progress and the results show that these scenarios can be an alternative description to the standard cosmological model. Certainly, much progress and physical properties for this dark coupling have yet to be explored, such as, the dark sector's particle mass scale dependence on CMB and LSS; more theoretical ground for the coupling function $Q$; finding the connection with gravitational waves, as well as others. Processes in this regard are still necessary and attention should be focused for more physical details of the dark coupling in future works.

\appendix
\begin{center} 
\squeezetable    
\begin{table*}                                                  
\begin{tabular}{ccccccccccccccc}                                
\hline\hline                                                                                                                    
Parameters & Planck 2018 & Planck 2018 + BAO & Planck 2018 + R19 \\ \hline

$\Omega_c h^2$ & $    0.1199_{-    0.0013-    0.0026}^{+    0.0013+    0.0027}$ & $    0.1199_{-    0.0013-    0.0026}^{+    0.0013+    0.0025}$ & $    0.1202_{-    0.0013-    0.0026}^{+    0.0013+    0.0025}$  \\

$\Omega_b h^2$ & $    0.02240_{-    0.00015-    0.00030}^{+    0.00015+    0.00029}$ & $    0.02238_{-    0.00015-    0.00028}^{+    0.00015+    0.00029}$ & $    0.02237_{-    0.00017-    0.00028}^{+    0.00014+    0.00031}$  \\

$100\theta_{MC}$ & $    1.04093_{-    0.00031-    0.00060}^{+    0.00031+    0.00062}$ &  $    1.040947_{-    0.00031-    0.00062}^{+    0.00030+    0.00060}$ & $    1.04091_{-    0.00032-    0.00063}^{+    0.00033+    0.00063}$  \\

$\tau$ & $    0.054_{-    0.0082-    0.015}^{+    0.0074+    0.016}$ & $    0.055_{-    0.0076-    0.015}^{+    0.0077+    0.015}$  & $    0.054_{-    0.0079-    0.015}^{+    0.0069+    0.017}$ \\

$n_s$ & $    0.9654_{-    0.0043-    0.0087}^{+    0.0043+    0.0087}$ & $    0.9656_{-    0.0043-    0.0080}^{+    0.0041+    0.0087}$  & $    0.9648_{-    0.0041-    0.0081}^{+    0.0040+    0.0083}$ \\

${\rm{ln}}(10^{10} A_s)$ & $    3.044_{-    0.017-    0.031}^{+    0.015+    0.033}$ & $    3.045_{-    0.015-    0.032}^{+    0.015+    0.031}$ & $    3.044_{-    0.017-    0.030}^{+    0.015+    0.034}$  \\

$w_0$ & $   -1.585_{-    0.356-    0.415}^{+    0.152+    0.482}$ & $   -1.039_{-    0.055-    0.118}^{+    0.060+    0.117}$  & $   -1.231_{-    0.049-    0.107}^{+    0.051+    0.100}$ \\

$\Omega_{m0}$ & $    0.197_{-    0.058-    0.069}^{+    0.020+    0.098}$ & $    0.304_{-    0.012-    0.024}^{+    0.012+    0.025}$ & $    0.260_{-    0.0099-    0.019}^{+    0.010+    0.021}$ \\

$\sigma_8$ & $    0.973_{-    0.047-    0.145}^{+    0.097+    0.120}$ & $    0.822_{-    0.020-    0.038}^{+    0.020+    0.039}$ & $    0.876_{-    0.018-    0.033}^{+    0.017+    0.034}$  \\

$H_0$ & $   86.66_{-    5.90-   17.71}^{+   12.58+   14.95}$ & $   68.57_{-    1.55-    2.78}^{+    1.41+    3.01}$ & $   74.28_{-    1.39-    2.72}^{+    1.36+    2.87}$ \\

\hline\hline                                                                                                                    
\end{tabular}                                                                                                                   
\caption{Constraints at 68\% and 95\% CL on free and derived parameters of the $w$CDM scenario. The parameter $H_0$ is measured in the units of km/s/Mpc.}
\label{tab:wCDM1}                                                                                                   
\end{table*}                                                                                                                     
\end{center}                                                                                                                    
\begingroup                                                                                                                     
\squeezetable                                                                                                                   
\begin{center}                                                                                                                  
\begin{table*}                                                                                                                   
\begin{tabular}{cccccccc}                                                                                                            
\hline\hline                                                                                                                    
Parameters & Planck 2018  & Planck 2018 + BAO & Planck 2018 + R19\\ \hline

$\Omega_c h^2$ & $    0.1200_{-    0.0014-    0.0027}^{+    0.0014+    0.0027}$ & $    0.1199_{-    0.0012-    0.0025}^{+    0.0012+    0.0025}$ & $    0.1202_{-    0.0014-    0.0027}^{+    0.0014+    0.0027}$ \\

$\Omega_b h^2$ & $    0.02238_{-    0.00016-    0.00031}^{+    0.00016+    0.00031}$ & $    0.02238_{-    0.00015-    0.00029}^{+    0.00015+    0.00028}$ & $    0.02235_{-    0.00015-    0.00030}^{+    0.00015+    0.00029}$ \\

$100\theta_{MC}$ & $    1.04090_{-    0.00032-    0.00063}^{+    0.00032+    0.00063}$ & $    1.04095_{-    0.00031-    0.00060}^{+    0.00031+    0.00059}$ & $    1.04087_{-    0.00031-    0.00063}^{+    0.00032+    0.00060}$ \\

$\tau$ & $    0.054_{-    0.0082-    0.015}^{+    0.0075+    0.016}$ & $    0.054_{-    0.0077-    0.015}^{+    0.0076+    0.016}$  & $    0.054_{-    0.0081-    0.015}^{+    0.0072+    0.016}$ \\

$n_s$ & $    0.9652_{-    0.0044-    0.0087}^{+    0.0044+    0.0087}$ & $    0.9655_{-    0.0041-    0.0080}^{+    0.0041+    0.0081}$  &  $    0.9644_{-    0.0043-    0.0086}^{+    0.0043+    0.0086}$ \\

${\rm{ln}}(10^{10} A_s)$ & $    3.044_{-    0.016-    0.031}^{+    0.016+    0.032}$ & $    3.044_{-    0.016-    0.031}^{+    0.016+    0.031}$ & $    3.044_{-    0.017-    0.030}^{+    0.015+    0.033}$ \\

$w_0$ & $   -1.589_{-    0.368-    0.411}^{+    0.144+    0.507}$ & $   -1.044_{-    0.057-    0.136}^{+    0.074+    0.132}$ & $   -1.249_{-    0.056-    0.141}^{+    0.078+    0.134}$ \\

$\Omega_{m0}$ & $    0.201_{-    0.060-    0.072}^{+    0.020+    0.104}$ & $    0.303_{-    0.012-    0.025}^{+    0.012+    0.025}$ & $    0.261_{-    0.011-    0.020}^{+    0.010+    0.022}$  \\

$\sigma_8$ & $    0.962_{-    0.052-    0.153}^{+    0.100+    0.129}$ & $    0.823_{-    0.020-    0.039}^{+    0.020+    0.040}$ & $    0.870_{-    0.019-    0.043}^{+    0.023+    0.041}$ \\

$H_0$ & $   86.03_{-    5.88-   17.96}^{+   12.74+   15.15}$ & $   68.71_{-    1.63-    3.03}^{+    1.43+    3.10}$  & $   74.29_{-    1.42-    2.75}^{+    1.41+    2.80}$ \\

$M_{\nu}$ & $  < 0.287$ & $   < 0.184$  & $    < 0.296$ \\

\hline\hline                                                                                                                    
\end{tabular}                                                                                                                   
\caption{Same as in table \ref{tab:wCDM1}, but for the $w$CDM +  $M_{\nu}$ scenario. The bound on $M_\nu$ is reported at 95\% CL in the units of eV.}
\label{tab:wCDM2}                                                                                                   
\end{table*}                                                                                                                     
\end{center}                                                                                                                    
\endgroup                                                        \begingroup                                                    
\squeezetable                                                   
\begin{center}                                                  
\begin{table*}                                              
\begin{tabular}{ccccccccccccccc}                                                                                                            
\hline\hline                                                                                                                    
Parameters & Planck 2018  & Planck 2018 + BAO & Planck 2018 + R19 \\ \hline

$\Omega_c h^2$ & $    0.1180_{-    0.0030-    0.0058}^{+    0.0030+    0.0059}$ & $    0.1183_{-    0.0031-    0.0057}^{+    0.0030+    0.0060}$ & $    0.1178_{-    0.0031-    0.0058}^{+    0.0030+    0.0061}$ \\

$\Omega_b h^2$ & $    0.02225_{-    0.00022-    0.00044}^{+    0.00022+    0.00044}$ & $    0.02229_{-    0.00021-    0.00040}^{+    0.00021+    0.00041}$ & $    0.02222_{-    0.00022-    0.00044}^{+    0.00022+    0.00044}$ \\

$100\theta_{MC}$ & $    1.04114_{-    0.00045-    0.00087}^{+    0.00044+    0.00087}$ & $    1.04113_{-    0.00044-    0.00085}^{+    0.00043+    0.00085}$ & $    1.04114_{-    0.00045-    0.00086}^{+    0.00045+    0.00088}$ \\

$\tau$ & $    0.0532_{-    0.0074-    0.015}^{+    0.0073+    0.016}$ & $    0.054_{-    0.0077-    0.016}^{+    0.0077+    0.016}$ & $    0.053_{-    0.0075-    0.016}^{+    0.0076+    0.016}$ \\

$n_s$ & $    0.9594_{-    0.0084-    0.0164}^{+    0.0083+    0.0167}$ & $    0.9615_{-    0.0079-    0.0155}^{+    0.0079+    0.0159}$  & $    0.9585_{-    0.0085-    0.0171}^{+    0.0086+    0.0165}$ \\

${\rm{ln}}(10^{10} A_s)$ & $    3.036_{-    0.018-    0.036}^{+    0.018+    0.036}$ & $    3.039_{-    0.018-    0.036}^{+    0.018+    0.037}$ & $    3.036_{-    0.018-    0.037}^{+    0.018+    0.037}$ \\

$w_0$ & $   -1.590_{-    0.408-    0.410}^{+    0.120+    0.511}$ & $   -1.057_{-    0.062-    0.143}^{+    0.080+    0.136}$ & $   -1.291_{-    0.074-    0.189}^{+    0.103+    0.175}$  \\

$\Omega_{m0}$ & $    0.206_{-    0.064-    0.078}^{+    0.023+    0.107}$ & $    0.303_{-    0.013-    0.025}^{+    0.013+    0.025}$  & $    0.257_{-    0.012-    0.022}^{+    0.011+    0.024}$ \\

$\sigma_8$ & $    0.950_{-    0.057-    0.153}^{+    0.102+    0.136}$ &  $    0.821_{-    0.020-    0.041}^{+    0.020+    0.040}$   & $    0.870_{-    0.019-    0.046}^{+    0.024+    0.043}$  \\

$H_0$ & $   84.45_{-    6.85-   18.11}^{+   12.71+   16.16}$ & $   68.29_{-    1.77-    3.19}^{+    1.63+    3.43}$  & $   74.17_{-    1.38-    2.77}^{+    1.38+    2.75}$ \\

$M_{\nu}$ & $  < 0.314$ & $ < 0.193$  &  $ < 0.324$ \\

$N_{\rm eff}$ & $    2.90_{-    0.18-    0.36}^{+    0.19+    0.37}$ & $    2.94_{-    0.19-    0.35}^{+    0.18+    0.36}$ &  $    2.88_{-    0.19-    0.37}^{+    0.19+    0.37}$ \\

\hline\hline                                                                                                                    
\end{tabular}                                               
\caption{Same as in table \ref{tab:wCDM1}, but for the $w$CDM +  $M_{\nu}$ + $N_{\rm eff}$ scenario. The bound on $M_\nu$ is reported at 95\% CL in the units of eV.}
\label{tab:wCDM3}                                                                                                   
\end{table*}                                                
\end{center}                     
\section{Updated constraints on the $w$CDM cosmological model}
\label{appendix}

In this appendix, we present our results for the $w$CDM cosmological model. This scenario is characterized by a dynamic dark energy density with constant EoS, $w$, together with uncoupled dark matter, baryons, radiation. The model is actually a simple extension of the standard cosmological model $\Lambda$CDM by including the dark energy equation of state, $w$, that means, $\Lambda$CDM + $w$. The parameter space of the model is the following. 

\begin{eqnarray}
\mathcal{P} \equiv\Bigl\{\Omega_{b}h^2, \Omega_{c}h^2, 100\theta_{MC}, \tau, n_{s}, \log[10^{10}A_{s}], w \Bigr\}.
\label{PS_wcdm1}
\end{eqnarray}

When we consider the inclusion of neutrinos, specifically, $M_{\nu}$ and $N_{\rm eff}$  (the scenario is labeled as $w$CDM + $M_{\nu}$ + $N_{\rm eff}$, the full parametric space of the model scenario is given by:

\begin{eqnarray}
\mathcal{P} \equiv\Bigl\{\Omega_{b}h^2, \Omega_{c}h^2, 100\theta_{MC}, \tau, n_{s}, \log[10^{10}A_{s}], w,  \nonumber\\ 
M_{\nu}, N_{\rm eff} \Bigr\}.
\label{PS_wcdm2}
\end{eqnarray}

In Tables \ref{tab:wCDM1}, \ref{tab:wCDM2} and \ref{tab:wCDM3}, we summarize the main results of the statistical analyses carried out for the observational datasets, namely, Planck 2018, Planck 2018  + BAO, and Planck 2018 + R19, considering three distinct scenarios, namely, $w$CDM, $w$CDM + $M_{\nu}$ and $w$CDM + $M_{\nu}$ + $N_{\rm eff}$. These results are interesting for a direct comparison between IVS and $w$CDM models, both in the presence of neutrinos properties, in order to check whether the underlying cosmological scenario may reveal any inherent features between $M_{\nu}$ and $N_{\rm eff}$. A discussion in this sense has been given in the main text of the article. Here, we only summarize briefly the results of the $w$CDM scenario.  

For Planck 2018  data only, we find $w < -1$ at 95\% CL, with a higher density of DE (consequently less DM) at late time, compared to the standard $\Lambda$CDM model, together with a high $H_0$ value. The $H_0$ constraint is high enough to be compatible with local measurements at 68\% CL. Thus, this simple extension of the $\Lambda$CDM model, with a preference for a phantom behavior on the EoS, is also able to alleviate the $H_0$ tension. We also analyze the model for the combined data, namely, Planck 2018 + BAO, Planck 2018 + R19, and found that for the latter combination, i.e., Planck 2018 + R19, $w < -1$ at 99\% CL. Thus, a strong preference for a phantom DE fluid at late time is reported. In light of the Planck 2018 + BAO analysis, we find that the EoS is consistent with $w = -1$ at 68\% CL. Therefore, in this joint analysis (Planck 2018 + BAO), we found no deviation from $\Lambda$CDM cosmology. It is important to remember that the BAO measurements are sometimes obtained adopting a fiducial cosmology (at $\Lambda$CDM level). Thus, the inclusion of BAO is motivated to break the degeneracy in the parametric space of some parameters. These results actually update the  constraints on the simple $w$CDM model in light of the final Planck 2018 release.

\begin{acknowledgments}
The authors thank the referee for several essential comments which helped the manuscript qualitatively.  WY was supported by the  National Natural Science Foundation of China under Grants No. 11705079 and No. 11647153. SP has been supported by the Mathematical Research Impact-Centric Support Scheme (MATRICS), File No. MTR/2018/000940, given by the Science and Engineering Research Board (SERB), Govt. of India. RCN would like to thank the agency FAPESP for financial support under the project \# 2018/18036-5. DFM thanks the Research Council of Norway for their support. Computations were performed on resources provided by UNINETT Sigma2 -- the National Infrastructure for High Performance Computing and Data Storage in Norway.  
\end{acknowledgments}


\end{document}